\documentclass[prd,superscriptaddress,amsfonts,amssymb,amsmath,showpacs,onecolumn]{revtex4-2}
\usepackage{bm}
\usepackage{amsfonts}
\usepackage{latexsym}
\usepackage[utf8]{inputenc}
\usepackage{graphicx}
\usepackage{amsmath}
\usepackage{palatino}
\usepackage{mathpazo}
\usepackage{textcomp}
\usepackage{float}
\usepackage{booktabs}
\usepackage{dcolumn}
\usepackage{ragged2e}
\usepackage{hyperref}
\hypersetup{colorlinks,citecolor=blue}
\hypersetup{colorlinks=true,linkcolor=blue,filecolor=magenta,urlcolor=blue}
\usepackage{amsmath}
\usepackage{subfigure}
\usepackage[]{natbib}
\usepackage{xcolor}
\usepackage{orcidlink}
\usepackage{epsfig}
\usepackage{caption}
\usepackage[toc]{appendix}
\usepackage{commath}
\usepackage{cancel}
\usepackage{csquotes}
\usepackage{placeins}

\usepackage{multirow}

\def\jnl@style{\it}
\def\aaref@jnl#1{{\jnl@style#1}}

\def\aaref@jnl#1{{\jnl@style#1}}

\def\aj{\aaref@jnl{AJ}}                   % Astronomical Journal
\def\apj{\aaref@jnl{ApJ}}                 % Astrophysical Journal
\def\apjl{\aaref@jnl{ApJ}}                % Astrophysical Journal, Letters
\def\apjs{\aaref@jnl{ApJS}}               % Astrophysical Journal, Supplement
\def\apss{\aaref@jnl{Ap\&SS}}             % Astrophysics and Space Science
\def\aap{\aaref@jnl{A\&A}}                % Astronomy and Astrophysics
\def\aapr{\aaref@jnl{A\&A~Rev.}}          % Astronomy and Astrophysics Reviews
\def\aaps{\aaref@jnl{A\&AS}}              % Astronomy and Astrophysics, Supplement
\def\mnras{\aaref@jnl{Mon.~Not.~Roy.~Astron.~Soc.}}             % Monthly Notices of the RAS
\def\prd{\aaref@jnl{Phys.~Rev.~D}}        % Physical Review D
\def\prc{\aaref@jnl{Phys.~Rev.~C}}  % Physical Review C
\def\prl{\aaref@jnl{Phys.~Rev.~Lett.}}    % Physical Review Letters
\def\qjras{\aaref@jnl{QJRAS}}             % Quarterly Journal of the RAS
\def\skytel{\aaref@jnl{S\&T}}             % Sky and Telescope
\def\ssr{\aaref@jnl{Space~Sci.~Rev.}}     % Space Science Reviews
\def\zap{\aaref@jnl{ZAp}}                 % Zeitschrift fuer Astrophysik
\def\nat{\aaref@jnl{Nature}}              % Nature
\def\aplett{\aaref@jnl{Astrophys.~Lett.}} % Astrophysics Letters
\def\apspr{\aaref@jnl{Astrophys.~Space~Phys.~Res.}} % Astrophysics Space Physics Research
\def\physrep{\aaref@jnl{Phys.~Rep.}}      % Physics Reports
\def\physscr{\aaref@jnl{Phys.~Scr}}       % Physica Scripta
\def\commat{\aaref@jnl{Comm.~Math.~Phys.}}              % Communications in Mathematical Physics
\def\science{\aaref@jnl{Science}}               % Science
\def\cqg{\aaref@jnl{Classical Quant.~Grav.}}            % Classical and Quantum Gravity
\def\jpcs{\aaref@jnl{JPCS}}                                     % Journal of Physics Conference Series
\def\ijmpd{\aaref@jnl{Int.~J.~Mod.~Phys.~D}}                    % International Journal of Modern Physics D
\def\grg{\aaref@jnl{Gen.~Relat.~Gravit.}}               % General Relativity and Gravitation
\def\rpp{\aaref@jnl{Rep.~Prog.~Phys.}}          % Reports on Progress in Physics
\def\npa{\aaref@jnl{Nucl.~Phys.~A}}        % Nuclear Physics A
\def\lrr{\aaref@jnl{Living Rev.~Rel.}}                   % Living reviews in relativity
\def\jcap{\aaref@jnl{J.~Cosmology Astropart.~Phys.}}    % Journal of cosmology and astroparticle physics
\def\rmp{\aaref@jnl{Rev.~Mod.~Phys.}}   %Reviews of modern physics
\def\epjc{\aaref@jnl{Eur.~Phys.~J.~C}}
\def\plb{\aaref@jnl{~Phy.~Lett.~B}}
\def\mpla{\aaref@jnl{Mod.~Phy.~Lett.~A}}
\def\arxiv{\aaref@jnl{arxiv.org}}

\allowdisplaybreaks[1]
\begin{document}

\color{black}

\title{Observational analysis of late-time acceleration in $f(Q, L_m)$ gravity}

\author{Kairat Myrzakulov\orcidlink{0000-0002-4189-8596}}\email[Email: ]{krmyrzakulov@gmail.com} 
\affiliation{Department of General \& Theoretical Physics, L.N. Gumilyov Eurasian National University, Astana, 010008, Kazakhstan.}

\author{M. Koussour\orcidlink{0000-0002-4188-0572}}
\email[Email: ]{pr.mouhssine@gmail.com}
\affiliation{Department of Physics, University of Hassan II Casablanca, Morocco.} 

\author{O. Donmez\orcidlink{0000-0001-9017-2452}}
\email[Email: ]{orhan.donmez@aum.edu.kw}
\affiliation{College of Engineering and Technology, American University of the Middle East, Egaila 54200, Kuwait.}

\author{A. Cilli}
\email[Email: ]{acilli@yildiz.edu.tr} 
\affiliation{Yildiz Technical University, Department of Physics, Istanbul Turkey.}

\author{E. G\"udekli}
\email[Email: ]{gudekli@istanbul.edu.tr} 
\affiliation{Department of Physics, Istanbul University, Istanbul 34134, Turkey.}

\author{J. Rayimbaev\orcidlink{0000-0001-9293-1838}}
\email[Email: ]{javlon@astrin.uz}
\affiliation{Institute of Fundamental and Applied Research, National Research University TIIAME, Kori Niyoziy 39, Tashkent 100000, Uzbekistan.}
\affiliation{University of Tashkent for Applied Sciences, Str. Gavhar 1, Tashkent 100149, Uzbekistan.}
\affiliation{Urgench State University, Kh. Alimjan Str. 14, Urgench 221100, Uzbekistan}
\affiliation{Shahrisabz State Pedagogical Institute, Shahrisabz Str. 10, Shahrisabz 181301, Uzbekistan.}
%\date{\today}

\begin{abstract}
In this study, we explored late-time cosmology within an extended class of theories based on $f(Q, L_m)$ gravity. This theory generalizes $f(Q)$ gravity by incorporating a non-minimal coupling between the non-metricity $Q$ and the matter Lagrangian $L_m$, analogous to the $f(Q,T)$ theory. The coupling between $Q$ and $L_m$ leads to the non-conservation of the matter energy-momentum tensor. We first investigated a cosmological model defined by the functional form $f(Q, L_m) = \alpha Q + \beta L_m^n$, where $\alpha$, $\beta$, and $n$ are constants. The derived Hubble parameter $H(z) = H_0 (1+z)^{\frac{3n}{2(2n-1)}}$ indicates that $n$ significantly influences the scaling of $H(z)$ over cosmic history, with $n > 2$ suggesting accelerated expansion. We also examined the simplified case of $n = 1$, leading to the linear form $f(Q, L_m) = \alpha Q + \beta L_m$, consistent with a universe dominated by non-relativistic matter. Using various observational datasets, including $H(z)$ and Pantheon, we constrained the model parameters. Our analysis showed that the $f(Q, L_m)$ model aligns well with observational results and exhibits similar behavior to the $\Lambda$CDM model. The results, with $q_0 = -0.22 \pm 0.01$ across all datasets, indicate an accelerating universe, highlighting the model's potential as an alternative to $\Lambda$CDM.

\end{abstract}

\maketitle

\tableofcontents

\section{Introduction}\label{sec1}

Unmistakably, recent Type Ia supernovae (SNe Ia) observations have demonstrated that the universe is undergoing accelerated expansion \cite{Riess/1998,Riess/2004,Perlmutter/1999}. In the past twenty years, extensive observational evidence, such as analyses of large-scale structure (LSS) \cite{T.Koivisto,S.F.}, data from the Wilkinson Microwave Anisotropy Probe (WMAP) \cite{Spergel}, observations of cosmic microwave background (CMB) \cite{R.R.,Z.Y.}, and investigations into baryonic acoustic oscillations (BAO) \cite{D.J.,W.J.}, has consistently corroborated the observed acceleration of cosmic expansion. This accelerated expansion challenges earlier models and theories, leading physicists to propose the existence of dark energy (DE). DE is posited as a unique form of energy with significant negative pressure, comprising approximately 70\% of the universe's total energy and matter content. The most straightforward solution to explain DE involves introducing a cosmological constant (CC), which effectively explains the observed accelerated expansion and forms the foundation of the $\Lambda$CDM model, widely recognized for its remarkable success \cite{Zlatev/1999}. While observational evidence supports the $\Lambda$CDM model, it grapples with notable challenges. Chief among them is the discrepancy between predicted and observed values of the CC. Additionally, the cosmic coincidence problem posits that our universe exhibits nearly equal densities of matter and DE, which is unexpected. These issues drive researchers to seek alternative models that can account for the accelerated expansion of the universe more effectively. \cite{Weinberg/1989,Padmanabhan/2003,Steinhardt/1999}.

Recently, modified gravity theories have emerged as a prominent field within modern cosmology, seeking to provide a unified framework that explains both the early universe dynamics and the observed accelerated expansion in later stages. These theories extend Einstein's general relativity (GR) by modifying the Einstein-Hilbert action, thereby accommodating cosmic acceleration. Several modified gravity theories have been developed to explain the universe's acceleration across different cosmic epochs. Among these, $f(R)$ gravity theory, first introduced in \cite{Buchdahl/1970}, stands out as a fundamental and extensively investigated modification of GR. Many researchers have delved into different facets of $f(R)$ gravity, investigating its capability to induce both cosmic inflation and acceleration \cite{Dunsby/2010,Carroll/2004}. Another extension of the Einstein-Hilbert action involves a non-minimal coupling between matter and geometry, resulting in the $f(R, L_m)$ gravity theory. Harko and Lobo \cite{Harko/2010} proposed $f(R, L_m)$ gravity, where the gravitational Lagrangian is described as an arbitrary function involving both the Ricci scalar $R$ and the matter Lagrangian $L_m$. This theory has been studied extensively for its implications in astrophysics and cosmology \cite{Wang/2012,Goncalves/2023}. Myrzakulova et al. \cite{Myrzakulova/2024} investigated the dark energy phenomenon using $f(R, L_m)$ cosmological models, incorporating observational constraints. Myrzakulov et al. \cite{Myrzakulov/2024} conducted a study on the linear redshift parametrization of the deceleration parameter in the context of $f(R, L_m)$ gravity. In the same way, various other modified gravity theories with diverse cosmological implications have been proposed. These include the following: $f(G)$ theory \cite{Felice/2009,Bamba/2017,Goheer/2009}, $f(R, T)$ theory \cite{Harko/2011,Koussour_1/2022,Koussour_2/2022,Myrzakulov/2023,KK1}, $f(\mathcal{T}, B)$ theory \cite{Bahamonde/2018}, and more.

Within the geometrical framework, curvature is not the only fundamental geometric property; torsion and non-metricity are also essential elements associated with the connection in a metric space. GR has three equivalent representations. The first is the curvature-based representation, where torsion and non-metricity are zero. The second is the teleparallel equivalent of GR (TEGR) or $f(\mathcal{T})$ theory, which is entirely based on torsion $\mathcal{T}$ \cite{Ferraro/2007,Myrzakulov/2011,Capozziello/2011}. The third is the symmetric teleparallel equivalent of GR (STEGR) or $f(Q)$ theory, where non-metricity $Q$ is associated with gravity \cite{Jimenez/2018}. Weyl proposed an extension of Riemannian geometry following GR in an attempt to unify gravity and electromagnetism \cite{Weyl/1918}. In this framework, the non-metricity of spacetime generates the electromagnetic field, causing both the orientation and length of vectors to change under parallel transport. Though recently proposed, $f(Q)$ gravity theory has already demonstrated several intriguing and valuable applications in the literature. References \cite{Jimenez/2020,Khyllep/2021} present the initial cosmological solutions in $f(Q)$ gravity, while \cite{MK1, MK2, MK3, MK4, MK5, MK6, MK7, MK8} focus on topics such as cosmic acceleration and DE. Furthermore, Xu et al. \cite{Xu/2019,Xu/2020} proposed an extension of $f(Q)$ theory called $f(Q, T)$, in which the non-metricity scalar $Q$ is non-minimally coupled to the trace $T$ of the energy-momentum tensor. The cosmological implications of $f(Q, T)$ theory were investigated in preliminary studies by Xu et al. \cite{Xu/2019}. Subsequent research \cite{K6, K7} concentrated on examining late-time accelerated expansion within the framework of observational constraints. Further, various fields such as baryogenesis mechanisms \cite{Bhattacharjee}, inflationary scenarios \cite{Shiravand}, and perturbation analyses \cite{Najera} have garnered considerable attention. However, comparatively little research has been directed towards investigating the astrophysical implications of $f(Q, T)$ theory \cite{Tayde1, Sneha2, Bourakadi,KK2,KK3}. Expanding on the foundations laid by $f(R, L_m)$ theory, Hazarika et al. \cite{Hazarika/2024} augmented the STEGR by incorporating the matter Lagrangian into the Lagrangian density of $f(Q)$ theory, thereby introducing the $f(Q, L_m)$ framework. In this framework, both minimal and non-minimal couplings between geometry and matter are permissible. The authors obtained the general system of field equations by varying the action with respect to the metric tensor. They also examined the non-conservation of the matter energy-momentum tensor within this theory. In addition, they studied the cosmic evolution using a flat Friedmann-Lema\^{i}tre-Robertson-Walker (FLRW) metric, deriving the generalized Friedmann equations. Furthermore, the authors investigated two specific gravitational models characterized by different forms of the function $f(Q, L_m)$: $f(Q, L_m) = -\alpha Q + 2 L_m + \beta$ and $f(Q, L_m) = -\alpha Q + (2 L_m)^2 + \beta$ \cite{Hazarika/2024}. Myrzakulov et al. \cite{fQL1} explored the impact of bulk viscosity on late-time cosmic acceleration within an extended framework of $f(Q, L_m)$ gravity. They introduced a linear function $f(Q, L_m) = \alpha Q + \beta L_m$, deriving exact solutions under non-relativistic matter domination. By employing observational datasets, they investigated the cosmological implications of their model, aiming to reconcile theoretical predictions with observational constraints. In another work, Myrzakulov et al. \cite{fQL2} investigated late-time cosmology within the framework of $f(Q, L_m)$ gravity, focusing on the implications for observational cosmology. The authors derived analytical solutions for the field equations governing a flat FLRW universe and analyzed these solutions to explore the dynamics of the universe in this modified gravity theory.

In this study, we focus on a novel extension of $f(Q)$ gravity, namely $f(Q, L_m)$ gravity. We investigate the implications of $f(Q, L_m)$ gravity for cosmological models, particularly its ability to reproduce the observed accelerated expansion and its compatibility with observational data. We analyze the dynamics of $f(Q, L_m)$ gravity in comparison to the $\Lambda$CDM model, highlighting its strengths and implications for understanding the nature of DE. The manuscript is organized as follows: Sec. \ref{sec2} provides an introduction to $f(Q, L_m)$ gravity. Sec. \ref{sec3} outlines the cosmological model used in this study, detailing its model parameters and derived physical quantities. Sec. \ref{sec4} discusses a specific case within the model. Sec. \ref{sec5} focuses on constraining the model parameters using Hubble $H(z)$ and Pantheon datasets, followed by the conclusions in Sec. \ref{sec6}.

\section{Overview of $f(Q,L_{m})$ gravity}\label{sec2}

The action in $f(Q,L_{m})$ gravity is expressed as \cite{Hazarika/2024}
\begin{equation}
    S=\int f(Q,L_m) \sqrt{-g} d^4x, \label{Action}
\end{equation}
Here, $\sqrt{-g}$ represents the determinant of the metric, while $f(Q, L_m)$ denotes an arbitrary function of the non-metricity scalar $Q$ and the matter Lagrangian $L_m$.

The affine connection $Y^\alpha_{\;\;\mu\nu}$ in Weyl–Cartan geometry can be separated into three independent components, namely: the symmetric Levi-Civita connection $\Gamma^\alpha_{\;\;\mu\nu}$, the contortion tensor $K^\alpha_{\;\;\mu\nu}$, and the disformation tensor $L^\alpha_{\;\;\mu\nu}$. Therefore, it can be written as follows \cite{Xu/2019}:
\begin{equation}
Y^\alpha_{\;\;\mu\nu}=\Gamma^\alpha_{\;\;\mu\nu}+K^\alpha_{\;\;\mu\nu}+L^\alpha_{\;\;\mu\nu}.
\end{equation}

The Levi-Civita connection $\Gamma^\alpha_{\;\;\mu\nu}$ describes curvature and parallel transport in a torsion-free, metric-compatible space. It is defined solely by the metric $g_{\mu\nu}$ and its first derivatives, encapsulating the standard gravitational effects in GR,
\begin{equation}
    \Gamma^\alpha_{\;\;\mu\nu}=\frac12 g^{\alpha\lambda}(\partial_\mu g_{\lambda \nu}+\partial_\nu g_{\lambda \mu} - \partial_\lambda g_{\mu\nu}).
\end{equation}

The contortion tensor $K^\alpha_{\;\;\mu\nu}$ introduces torsion tensor $T^\alpha_{\;\;\mu\nu}$ into the geometry. It quantifies the deviation from a symmetric connection and represents the influence of intrinsic spin or angular momentum of matter on the spacetime structure. This tensor modifies the geodesics, allowing for the possibility of paths that are not purely determined by the metric,
\begin{equation}
    K^\alpha_{\;\;\mu\nu}=\frac{1}{2}(T^\alpha_{\;\;\mu\nu}+ T_{\mu\;\;\nu}^{\;\;\alpha}+T_{\nu\;\;\mu}^{\;\;\alpha}).
\end{equation}

The disformation tensor $L^\alpha_{\;\;\mu\nu}$ reflects the non-metricity of the connection, indicating that the length of vectors may change under parallel transport. This component is crucial in theories where the metric is not preserved during displacement, leading to more general geometric structures that extend beyond the Riemannian framework,
\begin{equation}
    L^\alpha_{\;\;\mu\nu}=\frac{1}{2}(Q^\alpha_{\;\;\mu\nu}-Q^{\;\;\alpha}_{\mu\;\;\nu}-Q^{\;\;\alpha}_{\nu\;\;\mu}).
\end{equation}

Regarding the Weyl–Cartan connection $Y^\alpha_{\;\;\mu\nu}$, the non-metricity tensor $Q_{\alpha\mu\nu}$ is the covariant derivative of the metric tensor. This relationship is expressed by $Q_{\alpha\mu\nu}=\nabla_\alpha g_{\mu\nu}$, and can be derived as follows:
\begin{equation}
    Q_{\alpha\mu\nu}= \partial_\alpha g_{\mu\nu} - Y^\beta_{\;\;\alpha\mu}g_{\beta\nu}-Y^\beta_{\;\;\alpha\nu}g_{\mu\beta}.
\end{equation}

Furthermore, we introduce the superpotential $P^\alpha_{\;\;\mu\nu}$, which is the non-metricity conjugate, as follows:
\begin{equation}
    P^\alpha_{\;\;\mu\nu}= -\frac{1}{2}L^\alpha_{\;\;\mu\nu}+\frac{1}{4}(Q^\alpha-\Tilde{Q}^\alpha)g_{\mu\nu}-\frac{1}{4}\delta^\alpha_{\;\;(\mu}Q_{\nu)},
\end{equation}
where $Q^\alpha=Q^{\alpha\;\;\mu}_{\;\;\mu}$ and $\Tilde{Q}^\alpha=Q_{\mu}^{\;\;\alpha\mu}$ are the non-metricity vectors. By contracting the superpotential tensor with the non-metricity tensor, one can derive the non-metricity scalar:
\begin{equation}
    Q=-Q_{\lambda\mu\nu}P^{\lambda\mu\nu}.
\end{equation}

By varying the gravitational action with respect to the metric tensor, the gravitational field equations can be derived,
\begin{equation}
\frac{2}{\sqrt{-g}}\nabla_\alpha(f_Q\sqrt{-g}P^\alpha_{\;\;\mu\nu}) +f_Q(P_{\mu\alpha\beta}Q_\nu^{\;\;\alpha\beta}-2Q^{\alpha\beta}_{\;\;\;\mu}P_{\alpha\beta\nu})+\frac{1}{2}f g_{\mu\nu}=\frac{1}{2}f_{L_m}(g_{\mu\nu}L_m-T_{\mu\nu}),\label{field}
\end{equation}
where $f_Q=\partial f(Q,L_m)/\partial Q$ and $f_{L_m}=\partial f(Q,L_m)/\partial L_m$. For the specific form $f(Q, L_m) = f(Q) + 2\,L_m $, this simplifies to the field equation of $f(Q)$ gravity \cite{Jimenez/2018}. Further, the matter's energy-momentum tensor $T_{\mu\nu}$ can be written as 
\begin{equation}
    T_{\mu\nu}=-\frac{2}{\sqrt{-g}}\frac{\delta(\sqrt{-g}L_m)}{\delta g^{\mu\nu}}=g_{\mu\nu}L_m-2\frac{\partial L_m}{\partial g^{\mu\nu}},
\end{equation}

Once more, varying the gravitational action with respect to the connection yields the field equations,
\begin{equation}
    \nabla_\mu\nabla_\nu\Bigl( 4\sqrt{-g}\,f_Q\,P^{\mu\nu}_{\;\;\;\;\alpha}+H_\alpha^{\;\;\mu\nu}\Bigl)=0,
\end{equation}
where $H_\alpha^{\;\;\mu\nu}$ is the hypermomentum density, which is defined as $H_\alpha^{\;\;\mu\nu}=\sqrt{-g}f_{L_m}\frac{\delta L_m}{\delta Y^\alpha_{\;\;\mu\nu}}$. Also, applying the covariant derivative to the field equation (\ref{field}) allows one to derive,
\begin{equation}
D_\mu\,T^\mu_{\;\;\nu}= \frac{1}{f_{L_m}}\Bigl[ \frac{2}{\sqrt{-g}}\nabla_\alpha\nabla_\mu H_\nu^{\;\;\alpha\mu} + \nabla_\mu\,A^{\mu}_{\;\;\nu} - \nabla_\mu \bigr( \frac{1}{\sqrt{-g}}\nabla_\alpha H_\nu^{\;\;\alpha\mu}\bigr) \Bigr]=B_\nu \neq 0.
\end{equation}

Therefore, the matter energy-momentum tensor is not conserved in $f(Q, L_m)$ gravity theory. The dynamical variables $Q$, $L_m$, and the thermodynamical quantities of the system are dependent on the non-conservation term $B_\nu$. Physically, this non-conservation suggests the presence of an additional force acting on massive test particles, leading to non-geodesic motion. It also reflects the exchange of energy within a specified volume of the physical system. Furthermore, the non-zero right-hand side of the energy-momentum tensor implies energy transfer processes or particle production. Notably, the energy-momentum tensor becomes conserved in the absence of $f_{L_m}$ terms in the equation \cite{Jimenez/2018}.

Here, we assume the flat FLRW geometry framework to study the cosmic expansion in $f(Q, L_m)$ gravity. This model assumes two fundamental properties \cite{ryden/2003}: homogeneity and isotropy. Homogeneity means the universe appears uniform at any given time, indicating that its density and structure are consistent across all large scales, implying the absence of preferred locations within the universe. Isotropy means that observations from any point in the universe reveal consistent physical laws and conditions in all directions, suggesting that the universe exhibits the same appearance when viewed from different vantage points. In FLRW geometry, the spacetime interval is given by
\begin{equation}
\label{FLRW}
    ds^2=-dt^2+a^2(t)(dx^2+dy^2+dz^2),
\end{equation}
where $a(t)$ is the scale factor of the universe. Using the metric (\ref{FLRW}), the non-metricity scalar is given by $Q = 6 H^2$, where $H = \frac{\dot{a}}{a}$ is the Hubble parameter, representing the rate of expansion of the universe.

In addition, the energy-momentum tensor for a perfect fluid, which in cosmology is defined as a fluid that has uniform density and pressure with no shear stress or viscosity, is expressed as
\begin{equation}
    T_{\mu\nu}=(\rho+p)u_{\mu}u_{\nu}+pg_{\mu\nu},
\end{equation}
where $\rho$ is the energy density, $p$ is the pressure, and $u^\mu$ is the four-velocity of the fluid, with components $u^\mu = (1, 0, 0, 0)$. 

Thus, the modified Friedmann equations, describing a universe dominated by a perfect fluid as matter in $f(Q, L_m)$ gravity, are given by \cite{fQL1}
\begin{eqnarray}
\label{F1}
    && 3H^2 =\frac{1}{4f_Q}\bigr[ f - f_{L_m}(\rho + L_m) \bigl],\\
   && \dot{H} + 3H^2 + \frac{\dot{f_Q}}{f_Q}H=\frac{1}{4f_Q}\bigr[ f + f_{L_m}(p - L_m) \bigl]. \label{F2}
\end{eqnarray}

In particular, for the case where $f(Q, L_m) = f(Q) + 2\,L_m$, the Friedmann equations simplify to $f(Q)$ gravity. This further simplification leads to the STEGR.

\section{Cosmological model with $f(Q, L_m) = \alpha Q +\beta L_{m}^n$} \label{sec3}

In this study, we investigate a cosmological model defined by a functional form that incorporates a linear dependence on non-metricity and power-law behaviors of the matter content \cite{Xu/2019},
\begin{equation}
    f(Q, L_m) = \alpha Q +\beta L_{m}^n,
\end{equation}
where $\alpha$, $\beta$ and $n$ are constants. Thus, we readily determine that $f_Q = \alpha$, and $f_{L_m} = \beta  n L_m^{n-1}$. For this particular functional form, with $L_m = \rho$ \cite{Harko/2015}, the modified Friedmann equations (\ref{F1}) and (\ref{F2}) are derived as
\begin{eqnarray}
3 H^2+\frac{\beta }{2\alpha }(2 n-1) \rho ^n&=&0,\label{F11}\\
2 \Dot{H}+3 H^2-\frac{\beta }{2\alpha } \rho ^{n} (n \omega-n +1 )&=&0, \label{F22}
\end{eqnarray}
where $\omega$ is the equation of state (EoS) parameter defined as $\omega = \frac{p}{\rho}$. It plays a crucial role in determining the dynamics and evolution of cosmic structures such as dark energy, dark matter, and radiation. The value of $\omega$ categorizes different types of matter and energy. For instance, in standard cosmological models, dark energy is often described by an EoS parameter $\omega = -1$, indicating a constant energy density over time, whereas ordinary matter like non-relativistic material has $\omega = 0$ and radiation typically has $\omega = \frac{1}{3}$. With our focus on late-time acceleration, we consider the dominance of non-relativistic matter in the universe, specifically $\omega = 0$. From modified Friedmann equations (\ref{F11}) and (\ref{F22}), we derive
\begin{eqnarray}
\label{rho1}
\rho^n&=&\frac{6 \alpha H^2}{\beta (1-2n)},\\
\Dot{H}&-&\frac{ 3 n}{2(1-2 n)}H^2=0. \label{H1}
\end{eqnarray}

To estimate the magnitude of the non-conservation effect, we can examine the continuity equation, which in the presence of non-conservation is modified to:
\begin{equation}
\dot{\rho} + 3H(1 + \omega)\rho = B_0,
\end{equation}
where $B_0$ represents the temporal component of the non-conservation term $ B_\nu$. From the modified Friedmann equations (\ref{F11}) and (\ref{F22}), we can infer that the non-conservation term $B_0$ is proportional to the coupling parameters $\alpha$ and $\beta$, and it scales with $\rho^n$. Thus, the magnitude of the non-conservation effect is of the order of $B_0 \sim \frac{\beta}{\alpha} \rho^n$. This expression shows that the strength of the non-conservation effect is directly proportional to the ratio $ \frac{\beta}{\alpha}$, which characterizes the coupling in the theory, and depends on the power of the energy density $\rho^n$. Higher values of $\beta$, as well as larger powers of $n$, lead to stronger non-conservation effects.

In order to reconcile theoretical results with cosmological observations, we express all cosmological parameters in terms of redshift $z$, using the relationship $a(t) = \frac{1}{1+z}$ (with $a(t_0) = 1$ assumed). By substituting $\frac{1}{H} \frac{d}{dt} = \frac{d}{d \ln(a)}$ into Eq. \eqref{H1}, we derive a first-order differential equation for the Hubble parameter, which is expressed as
\begin{equation}
\frac{dH}{d \ln(a)}-\frac{ 3 n}{2(1-2 n)}H=0.
\label{H2}    
\end{equation}

Integrating Eq. (\ref{H2}), the Hubble parameter in terms of $z$ can be expressed as
\begin{equation}
\label{Hz}
H(z)=H_0 (1+z)^{\frac{3 n}{2 (2 n-1)}},
\end{equation}
where $H_0 = H(z=0)$ represents the current value of the Hubble parameter. This finding indicates that the model parameters $\alpha$ and $\beta$ do not affect $H(z)$. The power-law dependence on $(1+z)$, governed by the exponent $\frac{3n}{2(2n-1)}$, dictates the scaling of $H(z)$ over cosmic history, with $n$ influencing the shape of the expansion curve. This form is commonly used in cosmological models to explore scenarios such as modified gravity or dark energy theories \cite{Kumar/2012,Rani/2015}, where $n$ can be adjusted to align theoretical predictions with observed phenomena like SNe Ia and Hubble measurements.

To characterize whether the cosmological expansion is accelerating or decelerating, we introduce the deceleration parameter $q$, defined as $q = -1 - \frac{\dot{H}}{H^2}$. Then, using Eq. (\ref{Hz}), we derive
\begin{equation}
q(z)=\frac{2-n}{2(2 n-1)},
\end{equation}
which remains constant as expected, reflecting the power-law type expansion of the model. The universe is decelerating, or slowing down in its expansion if $q > 0$. For $q > 0$, the inequality $\frac{2-n}{2(2 n - 1)} > 0$ must be satisfied, which implies that $\frac{1}{2} < n < 2$. On the other hand, if $q < 0$, the universe is accelerating in its expansion, suggesting that it is expanding faster with time. For $q < 0$, the inequality $\frac{2-n}{2(2 n - 1)} < 0$ must be satisfied, which occurs when $n < \frac{1}{2}$ or $n > 2$. A universe expanding at a constant rate, without acceleration or deceleration, corresponds to $q = 0$, which is satisfied when $n = 2$. In cosmological terms, the condition $n > 2$ indicates accelerated expansion, aligning with observational data that suggest an accelerating universe, dominated by dark energy or similar components.

The energy density expression in Eq. (\ref{rho1}) can be rewritten as a function of the redshift $z$ in the following form:
\begin{equation}
\rho(z)=\left\{ \frac{6\alpha H_{0}^{2}}{\beta \left( 1-2n\right) }\right\} ^{%
\frac{1}{n}}\left( 1+z\right) ^{\frac{3}{\left( 2n-1\right) }}.   
\end{equation}

This expression indicates that the energy density evolves as a power of $(1+z)$, where the power depends on the parameter $n$. For $n = 1$, the energy density behaves like that of a matter-dominated universe, which scales as $\rho \sim (1+z)^3$. For other values of $n$, this scaling changes, reflecting how the specific model under consideration modifies the universe's expansion history. The parameter $\alpha$ controls the overall magnitude of $\rho(z)$, while $\beta$ and $n$ influence how $\rho(z)$ evolves with redshift. The sign of the term $1-2n$ is also critical, as it determines whether the energy density remains positive throughout cosmic history.

Now, we can compare both the deceleration parameter and the matter density with the predictions from the $\Lambda$CDM model: In the $\Lambda$CDM model, $q(z)$ evolves from a positive value (deceleration) at high redshift to a negative value (acceleration) at low redshift. In contrast, in our model, $q(z)$ is constant. For the matter density, the $\Lambda$CDM model predicts a matter-dominated phase at early times, followed by dark energy domination at late times. The evolution of matter density in our model depends on the value of $n$.

\section{Cosmological model with $f(Q, L_m) = \alpha Q +\beta L_{m}$} \label{sec4}

We observe that when $n = 1$, the situation simplifies to the linear form of the $f(Q, L_m)$ function, given by $f(Q, L_m) = \alpha Q +\beta L_{m}$, where $\alpha$ and $\beta$ are constants \cite{fQL1}. In this case, we have $f_Q = \alpha$ and $f_{L_m} = \beta$. For $n=1$, the modified Friedmann equations (\ref{F11}) and (\ref{F22}) are derived as follows:
\begin{eqnarray}
3 H^2+\frac{\beta }{2\alpha } \rho&=&0,\label{F111}\\
2 \Dot{H}+3 H^2-\frac{\beta }{2\alpha } p &=&0. \label{F222}
\end{eqnarray}

For zero pressure, the Hubble parameter in Eq. (\ref{Hz}) is expressed as $H(z)=H_0 (1+z)^{\frac{3}{2}}$. This expression indicates that the rate of expansion of the universe, as described by the Hubble parameter, scales with the redshift $z$. The factor $(1+z)^{\frac{3}{2}}$ reflects the influence of non-relativistic matter on the expansion rate, signifying that as the redshift increases, the expansion rate increases as well. This behavior is consistent with a universe dominated by non-relativistic matter, where the density of matter decreases as the universe expands, leading to a deceleration ($q>0$) of the expansion over time. However, in the context of the late-time universe, the presence of dark energy or similar components could alter this trend, causing an accelerated expansion ($q<0$). In this study, we examined the case of cosmic acceleration resulting from the $f(Q, L_m)$ modified gravity through the power law of the function $f(Q, L_m)$, which produces the acceleration phase for $n > 2$. This analysis highlights the significance of the parameter $n$ in driving the accelerated expansion of the universe, aligning with observational data suggesting an accelerating universe. The transition from deceleration to acceleration in the cosmic expansion underscores the dynamic nature of the universe and the role of modified gravity theories in explaining this phenomenon. In the next section, we will use a variety of observational measurements to constrain the model parameter $n$ and the Hubble constant $H_0$. This approach aims to develop a physically realistic cosmological model that is consistent with astrophysical evidence.

\section{Observational constraints} \label{sec5}

At this point, FLRW cosmology has been investigated in the context of $f(Q,L_m)$ gravity. It is crucial to emphasize the validation of parameter values for robust cosmological analysis. This section focuses on observational interpretations within the current cosmological context. We employ statistical techniques, specifically Markov Chain Monte Carlo (MCMC), employing standard Bayesian methods within the emcee Python environment \cite{emcee}, to constrain parameters such as $H_0$ and $n$. Moreover, by employing the chi-squared function $\chi^2$, the following probability function yields the best-fit values for the model parameters \cite{dat1,dat2}:
\begin{equation}
\mathcal{L}\propto \exp(-\chi^2 / 2).
\end{equation}

Here, we focus on two primary datasets: Hubble parameter measurements $H(z)$ and Pantheon SNe Ia data. Further, we apply priors on the parameters: $60.0 < H_0 < 80.0$ to encompass the range of possible Hubble constant values, and $2 < n < 20$ to explore the full range of the model parameter $n$ and to identify the accelerating regime. Below, we present the $\chi^2$ function for these datasets:

\subsection{$H(z)$ datasets}

The Hubble parameter $H(z)$ values are commonly derived using the differential age method of galaxies. By applying the relation $H(z) = -\frac{1}{1+z} \frac{dz}{dt}$, one may estimate the Hubble parameter at a given redshift $z$. The term $\frac{dz}{dt}$ can be obtained from observations of massive and slowly evolving galaxies, referred to as Cosmic Chronometers (CC). 

In our study, we employed 31 CC data points compiled in \cite{Hz_data}. These data points are crucial for our analysis as they provide a measure of $H(z)$ over various redshifts. It is important to note that the correlation between these data points was not explicitly considered in our analysis. The data points were treated as uncorrelated, which is a common approach in such analyses. To determine the best-fit values for the model parameters, we define the $chi^2$ function as follows:
\begin{equation}
\chi^{2}_{Hz} = \sum_{i=1}^{31} \frac{\left[H(H_0,n, z_{i})-
H_{obs}(z_{i})\right]^2}{\sigma(z_{i})^2},
\end{equation}
where, the index $i$ iterates through the 31 data points, each corresponding to a specific redshift $z_{i}$. Further, $H(H_0,\alpha, z_{i})$ denotes the predicted Hubble parameter at redshift $z_{i}$, $H_{obs}(z_{i})$ represents the observed Hubble parameter at redshift $z_{i}$, and $\sigma(z_{i})$ signifies the uncertainty associated with the observed value at that redshift.

\subsection{Pantheon datasets}

SNe Ia plays a crucial role in explaining the expanding universe. Notably, spectroscopically collected SNe Ia data from various surveys, including the SuperNova Legacy Survey (SNLS), Sloan Digital Sky Survey (SDSS), Hubble Space Telescope (HST) survey, and the Panoramic Survey Telescope and Rapid Response System (Pan-STARRS1) provide solid evidence supporting this concept. These supernovae serve as "standard candles" due to their consistent peak luminosity, allowing precise measurements of distances in the universe. The recent data sample of SNe Ia, known as the Pantheon dataset, comprises 1048 distance modulus measurements covering a redshift range of $0.01 \leq z \leq 2.3$ \cite{Scolnic/2018}. To determine the best-fit values for the model parameters, we conduct a comparative analysis of the theoretical and observational values of the distance moduli $\mu(z_k)$. Theoretically,
\begin{equation}
\mu^{th}(z_k)=\mu_0+5\,\text{log}_{10}(\mathcal{D}_L(z_k)),
\end{equation}
with the nuisance parameter
\begin{equation}
\mu_0= 25+5\, \text{log}_{10}\left(\dfrac{1}{H_0 Mpc}\right),
\end{equation}
and the luminosity distance
\begin{equation}
\mathcal{D}_L(z)=(1+z)\int\limits_0^z \dfrac{c}{H(\xi)}d\xi.
\end{equation}

Here, $c$ represents the speed of light. Now, the $\chi^2$ function for the Pantheon dataset, incorporating the covariance matrix $\mathcal{C}_{SNe}$, is given by
\begin{equation}
\chi^2_{SNe}(\mu_0,H_0,n)=\sum\limits_{k,l=1}^{1048}\bar{\mu}_k \left(\mathcal{C}_{SNe}^{-1}\right)_{kl}\bar{\mu}_l,
\end{equation}
where, $\bar{\mu}_k=\mu^{th}(z_k,H_0,n)-\mu^{ob}(z_k)$.

\subsection{$H(z)$+Pantheon datasets}

In the previous sections, we analyzed two types of datasets: $H(z)$ and Pantheon. Here, we consider the combination of these two datasets. Similarly, to determine the best-fit parameters for our model using the combined $H(z)$+Pantheon dataset, we use the $\chi^2$ function as defined below,
\begin{equation}
    \chi^{2}_{total} = \chi^{2}_{Hz} + \chi^{2}_{SNe}.
\end{equation}

\begin{table*}[ht]
\begin{ruledtabular}
		    \centering
		    \begin{tabular}{c c c c c}
				Dataset & $H_{0}$ & $n$ & $q_0$\\ 
				\hline
				$H(z)$ & $67.33\pm 0.62$ & $15.0\pm 1.0$ & $-0.22\pm 0.01$\\
				Pantheon & $66.44\pm 0.57$  & $15.07\pm 0.99$& $-0.22\pm0.01$\\
				$H(z)$+Pantheon & $66.93\pm 0.47$ & $15.0\pm 1.0$ & $-0.22\pm 0.01$\\
			\end{tabular}
		   \end{ruledtabular}
\caption{The table shows the best-fit values of the model parameters derived from the observational datasets.}
\label{tab}%
\end{table*}

\begin{figure}[h]
\centering
\includegraphics[width=18cm,height=5.5cm]{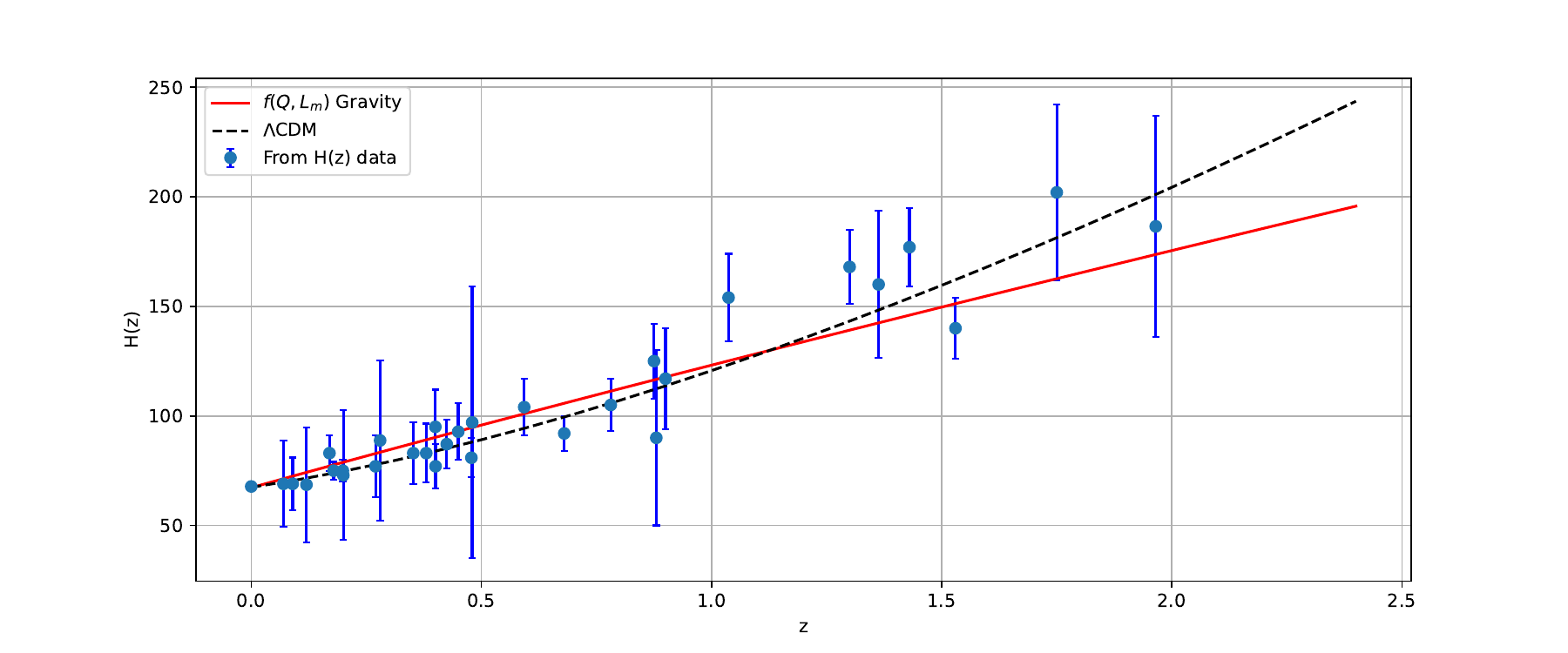}
\caption{The figure shows the error bars of $H(z)$ versus $z$ for our $f(Q,L_m)$ model. The solid red line represents the predicted curve for the $f(Q,L_m)$ model, while the black dotted line corresponds to the $\Lambda$CDM model. The blue dots indicate the 31 observed data points from the $H(z)$ datasets.}
\label{F_Hz}
\end{figure}

\begin{figure}[h]
\centering
\includegraphics[width=18cm,height=5.5cm]{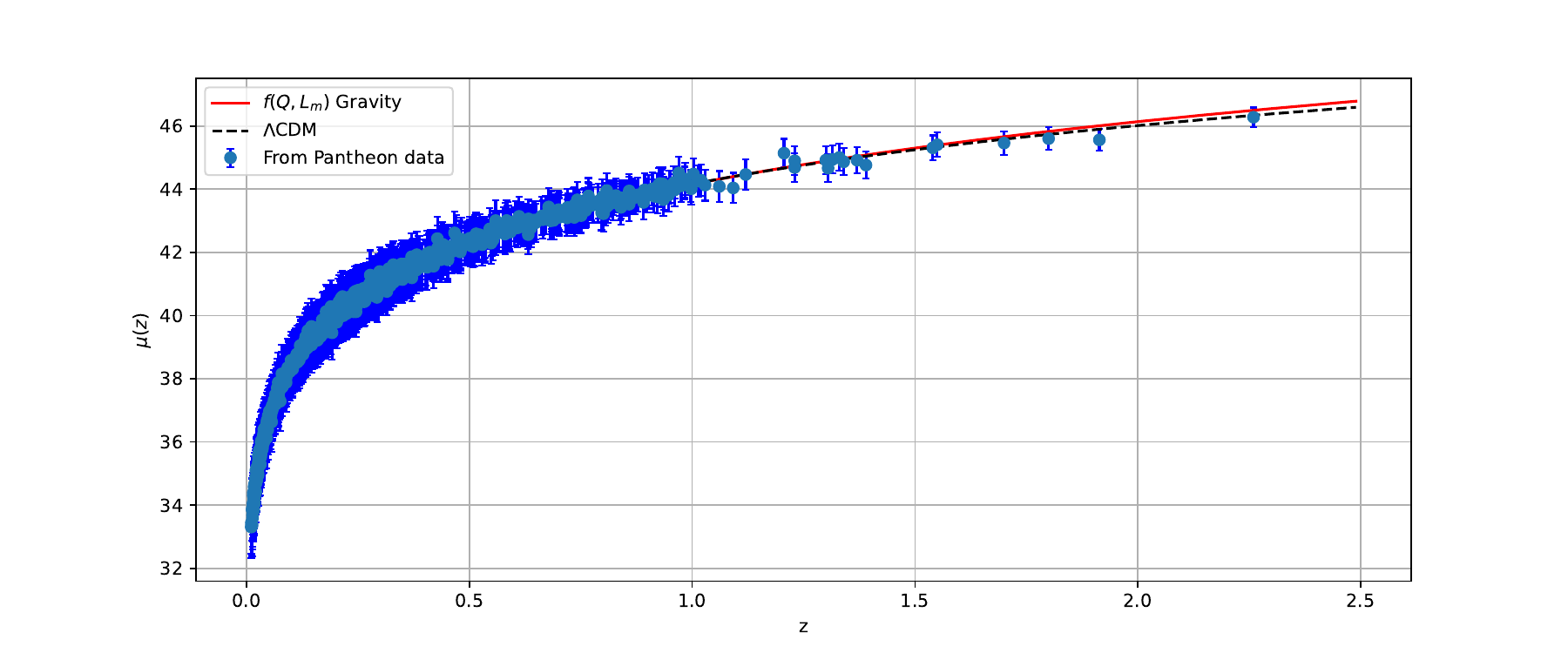}
\caption{The figure shows the error bars of $\mu(z)$ versus $z$ for our $f(Q,L_m)$ model. The solid red line represents the predicted curve for the $f(Q,L_m)$ model, while the black dotted line corresponds to the $\Lambda$CDM model. The blue dots indicate the 1048 observed data points from the Pantheon datasets.}
\label{F_mu}
\end{figure}

So far, we have examined various datasets and derived constrained values for the model parameters $H_0$ and $n$. Furthermore, we have generated two-dimensional likelihood contours displaying $1-\sigma$ and $2-\sigma$ errors corresponding to 68\% and 95\% confidence levels for the $H(z)$, Pantheon, and combined $H(z)$+Pantheon datasets. These contours are illustrated in Figs. \ref{F1} and \ref{F2}, respectively. From these figures, it is evident that the likelihood functions for all datasets closely resemble a Gaussian distribution. First, we analyzed the $H(z)$ dataset comprising 31 data points. For the Hubble constant $H_0$, we obtained a value of $67.33 \pm 0.62$. For the parameter $n$, which is crucial for producing the acceleration phase, the constrained value is $15.0 \pm 1.0$. This value falls within the accelerating range $n > 2$. Next, for the Pantheon dataset, which includes 1048 sample points, we obtained $H_0 = 66.44 \pm 0.57$ and $n = 15.07 \pm 0.99$. Finally, for the combined datasets, the values are $H_0 = 66.93 \pm 0.47$ and $n = 15.0 \pm 1.0$. To compare our model with the $\Lambda$CDM model, we examined the evolution of the Hubble parameter $H(z)$ and the distance modulus $\mu(z)$, using the constrained values of the model parameters $H_0$ and $n$ from the $H(z)$ and Pantheon samples. The results are illustrated in Figs. \ref{F_Hz} and \ref{F_mu}. For this purpose, we used the expression for the $\Lambda$CDM model: $H(z) = H_0 \sqrt{\Omega_0^m (1+z)^3 + (1-\Omega_0^m)}$, where $\Omega_0^m = 0.315\pm 0.007$ and $H_0 = 67.4 \pm 0.5$ $km.s^{-1}.Mpc^{-1}$ \cite{Planck/2014,Planck/2020}. It is observed that our $f(Q, L_m)$ model fits well with the observational results in both cases. In addition, our model's evolution is found to be quite similar to that of the $\Lambda$CDM model. Moreover, the deceleration parameter $q$ is well-known for its crucial role in describing the nature of the universe's expansion, whether it is accelerating ($q<0$) or decelerating ($q>0$). For the present model, we found $q_0 = -0.22 \pm 0.01 $ consistently across the $H(z)$, Pantheon, and combined $H(z)$ and Pantheon datasets. Our findings for $q_0$ are consistent with several observational studies (see References \cite{Hernandez,Basilakos,Roman,Jesus,Cunha}). It is noteworthy that our $f(Q, L_m)$ model consistently predicts an accelerating universe across all datasets analyzed. The summarized constraint values are presented in Tab. \ref{tab}.

\begin{figure}[h]
   \begin{minipage}{0.48\textwidth}
     \centering
     \includegraphics[width=1\linewidth]{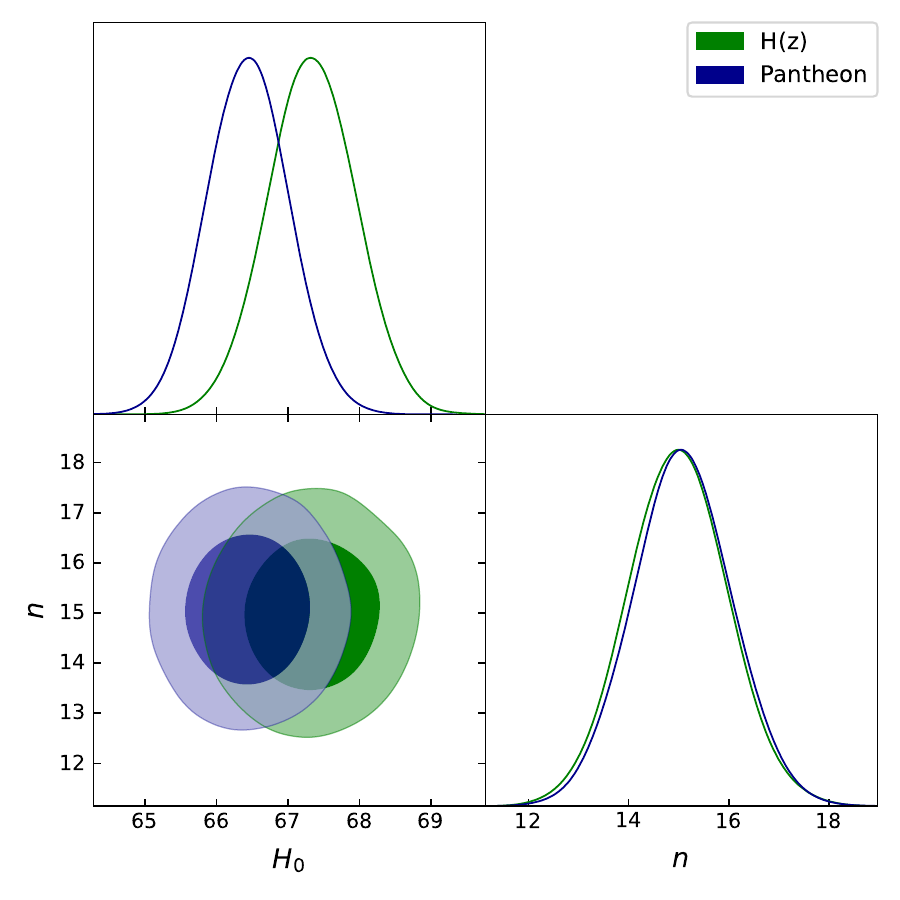}
     \caption{The figure shows $1-\sigma$ and $\sigma-2$ confidence regions for the parameters derived from the $H(z)$ and Pantheon datasets.}\label{F1}
   \end{minipage}\hfill
   \begin{minipage}{0.48\textwidth}
     \centering
     \includegraphics[width=1\linewidth]{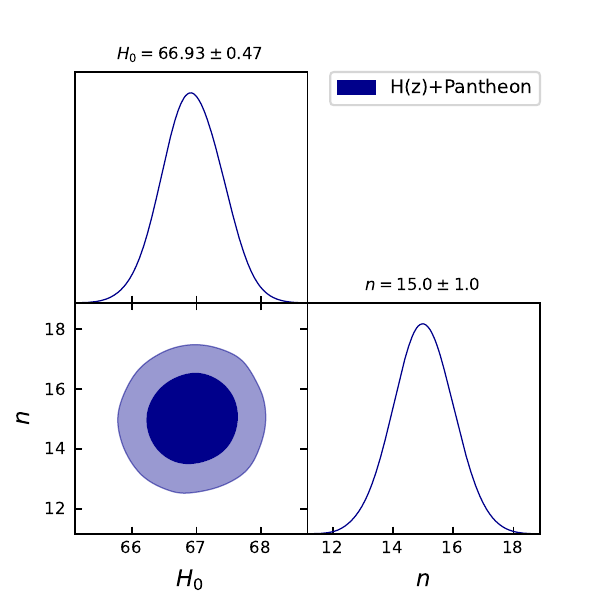}
     \caption{The figure shows $1-\sigma$ and $\sigma-2$ confidence regions for the parameters derived from the combined datasets.}\label{F2}
   \end{minipage}
\end{figure}

\section{Concluding remarks}\label{sec6}

In this study, we investigated late-time cosmology within an extended class of theories based on $f(Q, L_m)$ gravity \cite{Hazarika/2024}. This theory generalizes $f(Q)$ gravity by incorporating a non-minimal coupling between the non-metricity $Q$ and the matter Lagrangian $L_{m}$, similar to the $f(Q,T)$ theory where $T$ is replaced by $L_{m}$. The coupling between $Q$ and $L_{m}$ in the $f(Q, L_m)$ framework, akin to the trace-curvature couplings in energy-momentum tensor theories, leads to the non-conservation of the matter energy-momentum tensor. First, we investigated a cosmological model defined by the functional form $f(Q, L_m) = \alpha Q +\beta L_{m}^n$, where $\alpha$, $\beta$, and $n$ are constants. This model incorporates a linear dependence on non-metricity and a power-law behavior of the matter content. The Hubble parameter in terms of redshift $z$ was derived as $H(z) = H_0 (1+z)^{\frac{3n}{2(2n-1)}}$, indicating that the model parameters $\alpha$ and $\beta$ do not affect $H(z)$. The parameter $n$ dictates the scaling of $H(z)$ over cosmic history, with $n > 2$ indicating accelerated expansion, aligning with observational data.

We also examined the simplified case when $n = 1$, leading to the linear form $f(Q, L_m) = \alpha Q +\beta L_{m}$, and derived the corresponding modified Friedmann equations. For zero pressure, the Hubble parameter was expressed as $H(z) = H_0 (1+z)^{\frac{3}{2}}$, consistent with a universe dominated by non-relativistic matter. To constrain the model parameter $n$ and the Hubble constant $H_0$, we used various observational measurements, including $H(z)$ and Pantheon datasets. The analysis revealed that our $f(Q, L_m)$ model fits well with the observational results and shows a similar evolution to the $\Lambda$CDM model. Specifically, we obtained $H_0 = 67.33 \pm 0.62$ and $n = 15.0 \pm 1.0$ for the $H(z)$ dataset, $H_0 = 66.44 \pm 0.57$ and $n = 15.07 \pm 0.99$ for the Pantheon dataset, and $H_0 = 66.93 \pm 0.47$ and $n = 15.0 \pm 1.0$ for the combined datasets. The constant nature of the deceleration parameter $q_0 = -0.22 \pm 0.01$ in our model provides a unique perspective on cosmic acceleration. Unlike the $\Lambda$CDM model, which exhibits a transition redshift where $q(z)$ changes sign, our model shows a constant deceleration parameter, reflecting a different dynamic evolution. This behavior underscores the potential of our model to offer a novel explanation for the observed accelerated expansion of the universe. The results suggest that our $f(Q, L_m)$ model consistently predicts an accelerating universe and the constraint values are summarized in Tab. \ref{tab}. The model's ability to describe the accelerating expansion of the universe highlights its potential as an alternative to the $\Lambda$CDM model.

Future work should focus on further testing the $f(Q, L_m)$ model with additional observational data and exploring other cosmological probes. Investigating the model's behavior in different cosmological scenarios, such as during the early universe or in the context of perturbations, possibly offers further information. In addition, comparing our model with other modified gravity theories and exploring its implications for structure formation and cosmic evolution will be valuable.

\section*{Acknowledgment}
This research was funded by the Science Committee of the Ministry of Science and Higher Education of the Republic of Kazakhstan (Grant No. AP23487178).

\section*{Data Availability Statement}
This article does not introduce any new data.


\begin{thebibliography}{99}

\bibitem{Riess/1998} A.G. Riess et al., \textit{Astron. J.} \textbf{116} 1009 (1998).

\bibitem{Riess/2004} A.G. Riess et al., \textit{Astophys. J.} \textbf{607} 665-687 (2004).

\bibitem{Perlmutter/1999} S. Perlmutter et al., \textit{Astrophys. J.} \textbf{517} 377 (1999).

\bibitem{T.Koivisto} T. Koivisto, D.F. Mota, \textit{Phys. Rev. D} \textbf{73}, 083502 (2006).

\bibitem{S.F.} S.F. Daniel, \textit{Phys. Rev. D} \textbf{77}, 103513 (2008).

\bibitem{Spergel} D.N. Spergel et al., \textit{Astrophys.
J. Suppl.} \textbf{148}, 175 (2003).

\bibitem{R.R.} R.R. Caldwell, M. Doran, \textit{Phys. Rev. D} \textbf{69}, 103517 (2004).

\bibitem{Z.Y.} Z.Y. Huang et al., \textit{J. Cosm. Astrop. Phys.} \textbf{0605}, 013 (2006).

\bibitem{D.J.} D.J. Eisenstein et al., \textit{Astrophys. J.} \textbf{633}, 560 (2005).

\bibitem{W.J.} W.J. Percival at el., \textit{Mon. Not. R. Astron. Soc.} \textbf{401}, 2148 (2010).

\bibitem{Zlatev/1999} I. Zlatev, L. Wang, and P.J. Steinhardt, \textit{Phys. Rev. Lett.} \textbf{82} 896 (1999).

\bibitem{Weinberg/1989} S.Weinberg,  \textit{Rev. Mod. Phys.} \textbf{61} 1 (1989).

\bibitem{Padmanabhan/2003} T. Padmanabhan,  \textit{Phys. Rep.} \textbf{380} 235 (2003).

\bibitem{Steinhardt/1999} P.J. Steinhardt, L. Wang, and I. Zlatev, \textit{Phys. Rev. D} \textbf{59} 123504 (1999).

\bibitem{Buchdahl/1970} H.A. Buchdahl, \textit{Mon. Not. R. Astron. Soc.} \textbf{150} 1 (1970).

\bibitem{Dunsby/2010} P.K.S. Dunsby et al., \textit{Phys. Rev. D} \textbf{82} 023519 (2010).

\bibitem{Carroll/2004} S.M. Carroll et al., \textit{Phys. Rev. D} \textbf{70} 043528 (2004).

\bibitem{Harko/2010} T. Harko and F.S.N. Lobo, \textit{Eur. Phys. J. C} \textbf{70} 373–379 (2010).

\bibitem{Wang/2012} J. Wang and K. Liao, \textit{Class. Quantum Gravity} \textbf{29} 215016 (2012).

\bibitem{Goncalves/2023} B.S. Goncalves and P.H.R.S. Moraes, \textit{Fortschr. Phys.} \textbf{71} 2200153 (2023).

\bibitem{Myrzakulova/2024} S. Myrzakulova et al., \textit{Phys. Dark Universe} \textbf{43} 101399 (2024).

\bibitem{Myrzakulov/2024} Y. Myrzakulov et al., \textit{Phys. Dark Universe} \textbf{45} 101545 (2024).

\bibitem{Felice/2009} A. De Felice and S. Tsujikawa, \textit{Phys. Lett. B} \textbf{675} 1 (2009).

\bibitem{Bamba/2017} K. Bamba et al., \textit{Gen. Relativ. Gravit.} \textbf{49} 112 (2017).

\bibitem{Goheer/2009} N. Goheer et al., \textit{Phys. Rev. D} \textbf{79} 121301 (2009).

\bibitem{Harko/2011} T. Harko et al., \textit{Phys. Rev. D} \textbf{84} 024020 (2011).

\bibitem{Koussour_1/2022} M. Koussour and M. Bennai, \textit{Int. J. Geom. Methods Mod. Phys.} \textbf{19} 2250038 (2022).

\bibitem{Koussour_2/2022} M. Koussour and M. Bennai, \textit{Afr. Mat.} \textbf{33} 27 (2022).

\bibitem{Myrzakulov/2023} N. Myrzakulov et al., \textit{Chin. Phys. C} \textbf{47} 115107 (2023).

\bibitem{KK1} M. Koussour et al., \textit{Phys. Dark Universe} \textbf{46} 101577 (2024).

\bibitem{Bahamonde/2018} S. Bahamonde, M. Zubair, and G. Abbas, \textit{Phys. Dark Univ.} \textbf{19} 78-90 (2018).

\bibitem{Ferraro/2007} R. Ferraro and F. Fiorini, \textit{Phys. Rev. D} \textbf{75} 084031 (2007).

\bibitem{Myrzakulov/2011} R. Myrzakulov, \textit{Eur. Phys. J. C} \textbf{71} 1752 (2011).

\bibitem{Capozziello/2011} S. Capozziello et al., \textit{Phys. Rev. D} \textbf{84} 043527 (2011).

\bibitem{Jimenez/2018} J.B. Jimenez, L. Heisenberg, and T. Koivisto, \textit{Phys. Rev. D} \textbf{98} 044048 (2018).

\bibitem{Weyl/1918} H. Weyl, \textit{Sitzungsber. Preuss. Akad.Wiss.} \textbf{465} 1 (1918).

\bibitem{Jimenez/2020} J.B. Jimenez et al., \textit{Phys. Rev. D} \textbf{101} 103507 (2020).

\bibitem{Khyllep/2021} W. Khyllep et al., \textit{Phys. Rev. D} \textbf{103} 103521 (2021).

\bibitem{MK1} M. Koussour et al., \textit{Phys. Dark universe} \textbf{36},
101051 (2022).

\bibitem{MK2} M. Koussour and M. Bennai \textit{Chin. J. Phys. } \textbf{379},
339-347 (2022).

\bibitem{MK3} M. Koussour et al. \textit{Phys. Ann. Phys.} \textbf{445},
169092 (2022).

\bibitem{MK4} M. Koussour and A. De, \textit{Eur. Phys. J. C} \textbf{83}, 400 (2023).

\bibitem{MK5} M. Koussour et al., \textit{Fortschr. Phys.} \textbf{71}, 2200172 (2023).

\bibitem{MK6} M. Koussour et al., \textit{Nucl. Phys. B} \textbf{990}, 116158 (2023).

\bibitem{MK7} M. Koussour et al., \textit{J. High Energy Phys.} \textbf{37}, 15-24 (2023).

\bibitem{MK8} M. Koussour et al., \textit{J. High Energy Astrophys, } \textbf{35}, 43-51 (2022).

\bibitem{Xu/2019} Y. Xu et al., \textit{Eur. Phys. J. C} \textbf{79}, 708 (2019).

\bibitem{Xu/2020} Y. Xu et al., \textit{Eur. Phys. J. C} \textbf{80}, 449 (2020).

\bibitem{K6}  M. Koussour et al., \textit{Int. J. Mod. Phys. D} \textbf{31},  2250115 (2022).

\bibitem{K7}  M. Koussour et al., \textit{Chin. J. Phys.} \textbf{86},  300-312 (2023).

\bibitem{Bhattacharjee} S. Bhattacharjee and P. K. Sahoo, \textit{Eur. Phys. J. C} \textbf{80}, 289 (2020).

\bibitem{Shiravand} M. Shiravand, S. Fakhry, and M. Farhoudi, \textit{Phys. Dark Univ.} \textbf{37}, 101106 (2022).

\bibitem{Najera} A. Najera and A. Fajardo, \textit{J. Cosmo. Astro. Phys.} \textbf{2022}, 020 (2022).

\bibitem{Tayde1}  M. Tayde \textit{et al.}, \textit{Chinese Phys. C} \textbf{46}, 115101 (2022).

\bibitem{Sneha2} S. Pradhan, D. Mohanty, and P. K. Sahoo, \textit{Chinese Phys. C} \textbf{47}, 095104 (2023).

\bibitem{Bourakadi} K. El Bourakadi et al., \textit{Phys. Dark Universe} \textbf{41}, 101246 (2023).

\bibitem{KK2} M. Koussour et al., \textit{Phys. Dark Universe} \textbf{45}, 101527 (2024).

\bibitem{KK3} M. Koussour et al., \textit{Chin. J. Phys.} \textbf{90}, 108-120 (2024).

\bibitem{Hazarika/2024} A. Hazarika et al. \textit{arXiv}, arXiv:2407.00989 (2024).

\bibitem{fQL1} Y. Myrzakulov et al. \textit{arXiv}, arXiv:2407.08837 (2024).

\bibitem{fQL2} Y. Myrzakulov et al., \textit{Phys. Dark Universe} \textbf{46}, 101614 (2024).

\bibitem{ryden/2003} B. Ryden, \textit{ Introduction to Cosmology} (Addison Wesley, San Francisco, United States of America, 2003).

\bibitem{Harko/2015} T. Harko et al., \textit{Eur. Phys. J. C} \textbf{75}, 386 (2015).

\bibitem{Kumar/2012} S. Kumar, \textit{Mon. Not. Roy. Astron. Soc.} \textbf{422}, 2532 (2012).

\bibitem{Rani/2015} S. Rani et al., \textit{J. Cosmol. Astropart. Phys.} \textbf{03}, 031 (2015).

\bibitem{emcee} D. F. Mackey et al.,  \textit{Publ. Astron. Soc. Pac.}, \textbf{125}, 306 (2013).

\bibitem{dat1} Y. Myrzakulov et al.,  \textit{J. High Energy Astro. Phys.}, \textbf{43}, 209-216 (2024).

\bibitem{dat2} A. Errehymy et al.,  \textit{Phys. Dark Universe	}, \textbf{46}, 101555 (2024).

\bibitem{Hz_data} H. Yu, B. Ratra, and F.Y. Wang, \textit{Astrophys. J.}, \textbf{856}, 3 (2018).

\bibitem{Scolnic/2018} D.M. Scolnic et al., \textit{Astrophys. J.} \textbf{859}, 101(2018).

\bibitem{Planck/2014} Planck Collaboration XVI, \textit{Astron. Astrophys.}, \textbf{571}, A16 (2014). 

\bibitem{Planck/2020} N. Aghanim et al. (Planck), \textit{Astron. Astrophys.}, \textbf{641}, A6 (2020).

\bibitem{Hernandez} A. Hernandez-Almada et al., \textit{Eur. Phys. J. C}, \textbf{79}, 12 (2019).

\bibitem{Basilakos} S. Basilakos, F. Bauer, J. Sola, \textit{J. Cosmol. Astropart. Phys.}, \textbf{01}, 050 (2012).

\bibitem{Roman} J. Roman-Garza et al., \textit{Eur. Phys. J. C}, \textbf{79}, 890 (2019).

\bibitem{Jesus} J.F. Jesus et al., \textit{J. Cosmol. Astropart. Phys.}, \textbf{04}, 053 (2020).

\bibitem{Cunha} J.V. Cunha and J.A.S. Lima, \textit{Mon. Not. R. Astron. Soc.}, \textbf{390}, 210 (2008).

\end{thebibliography}
\end{document}